\documentclass[12pt]{article}
\usepackage{graphicx}  

\usepackage{amsmath}

\begin{document}

\title{Discrimination of Dark Matter Mass and Velocity Distribution by Directional Detection}
\author{Keiko I. Nagao{$^1$}\\
\small{$^1$Okayama University of Science, Okayama 700-0005, JAPAN}}
\date{}

\maketitle

\renewcommand{\thefootnote}{\fnsymbol{footnote}}
\footnote[0]{This work is based on the paper~\cite{Nagao:2017yil}.}

\begin{abstract}
Velocity distribution of dark matter is assumed to be isotropic in most cases, however, anisotropy is suggested in some simulations. 
Directional direct detection of dark matter is a hopeful way to discriminate the anisotropy of dark matter velocity distribution. 
We simulate the dark matter and target scattering in the directional direct detection, and investigate conditions required to discriminate the anisotropy. If dark matter mass is known, $O(10^3)-O(10^4)$ events are required for the discrimination if the dark matter mass is known by other experiments. We also  study the case that the dark matter mass is not known, and in analysis using both the recoil energy and the scattering angle data, both the dark matter mass and the anisotropy can be restricted much better than  the analysis only with either of them.
%\vspace{0.2cm}
\end{abstract}

%\begin{keyword}
\small{keywords : Dark Matter, Velocity Distribution, Direct Detection}
%\doi{10.2018/LHEP000001}
%\end{keyword}

\section{Introduction}
Weakly interacting massive particles (WIMPs) are one of a hopeful candidate of dark matter. 
In most cases of direct detection of WIMPs, its velocity distribution is supposed to be isotropic Maxwellian distribution for simplicity. 
By some numerical simulations, its anisotropy is suggested. 
In \cite{LNAT}, velocity distributions regarding each direction that agree with the simulation are derived.
Tangential velocity distribution $f(v_\phi)$, the radial one $f(v_r)$ and that of direction across the Galactic plane $f(v_z)$ in the Galactic rest frame are
\begin{align}
&f(v_\phi) = \frac{1-r}{N(v_{0,\mathrm{iso.}})} \exp\left[-v_\phi^2/v_{0,\mathrm{iso.}}^2\right] + 
		\frac{r}{N(v_{0,\mathrm{ani.}})} \exp\left[-(v_\phi-\mu)^2/v_{0,\mathrm{ani.}}^2\right]  \nonumber \\
&f(v_r) = \frac{1}{N(v_{0,r})} \exp\left[-v_r^2/v_{0,r}^2\right] \label{eq:velocitydistribution} \\
&f(v_z) = \frac{1}{N(v_{0,z})} \exp\left[-v_z^2/v_{0,z}^2\right]  \nonumber 
\end{align}
where $\mu=150$ km/s and the anisotropy parameter $r=0.25$. 
Note that the tangential direction corresponds to the direction of the Solar system's motion in the Galaxy, and the first and the second term are isotropic and anisotropic component, respectively. 

In directional direct detection of dark matter, not only the recoil energy of dark matter and target scattering but also the direction of the nuclear recoil can be detected.
Thus in the directional detection such anisotropy of the velocity distribution
can potentially be detected. Currently the projects are in the research and development~\cite{Mayet2016}. We study required conditions to discriminate the anisotropic distribution ($r=0.3$) from the conventional Maxwellian ($r=0$) in the directional direct detection in this study. 

\section{Numerical Simulation and Analysis}
\subsection{Numerical Simulation of directional direct detection}
We perform Monte-Carlo simulation of WIMP and target scattering and obtained the recoil energy $E_R$ and the scattering angle $\theta$ along the direction of dark matter wind to the Solar system. The procedure is as follows.
\begin{enumerate}
\item WIMPs with velocity following a velocity distribution $f(v)$ in the Galactic rest frame are generated. In this study, we suppose two kinds of the distribution, i.e., isotropic Maxwellian and the anisotropic distribution in Eq.~(\ref{eq:velocitydistribution}).
\item The WIMPs are converted to that with the velocity in the laboratory frame, and the WIMP and nucleus scattering in the directional direct detection is simulated. As the target atom, fluorine (F) as light target and  silver (Ag) as heavy target are supposed in the simulation. F is a typical target in gaseous directional detector, while Ag is contained in nuclear emulsion, a solid directional detector. 
\item Data of scattering angle and recoil energy are obtained. 
\end{enumerate}
By repeating above procedure, two datasets are generated. The first dataset is ideal data which has O($10^8$) event number, and the second dataset, called as ``pseudo-experimental'' dataset, has smaller number of event than the first one. By analyzing the similarity between the ideal and pseudo-experimental distributions, we try to discriminate the anisotropy of the velocity distribution. 

The analysis of the data depends on detector. Energy - angular distribution can be ideally obtained, while only directional data can be analyzed if the energy resolution of the detector is not good enough. Therefore both the energy - angular distribution and the angular histogram are studied. 

\subsection{Case 1 : Dark matter mass is known.}
\label{subsec:case1}
\begin{figure}[t!]
%\vspace{-0.5cm}
 \centering
  \includegraphics[width=5.5cm,clip]{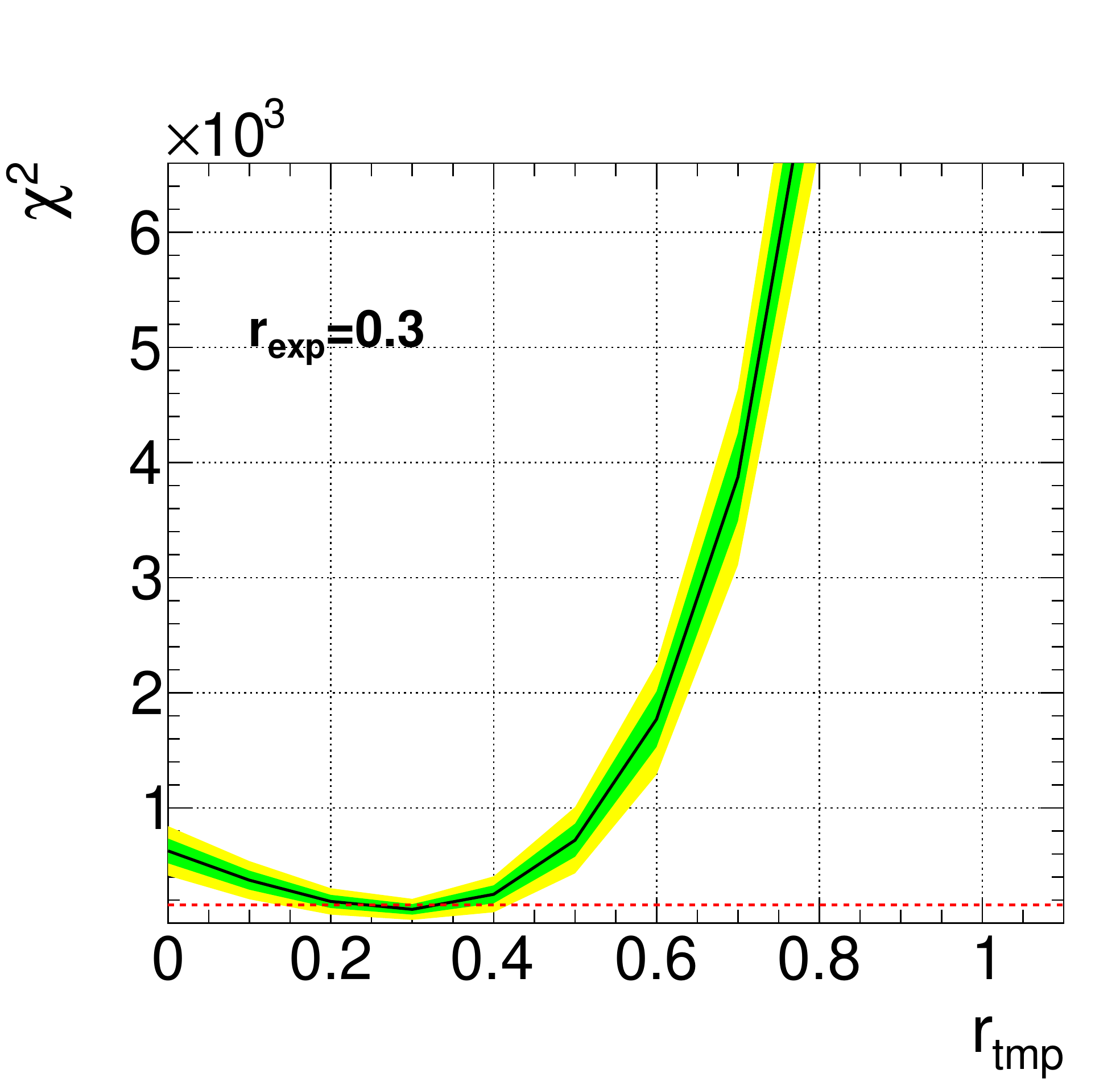}
  \includegraphics[width=5.5cm,clip]{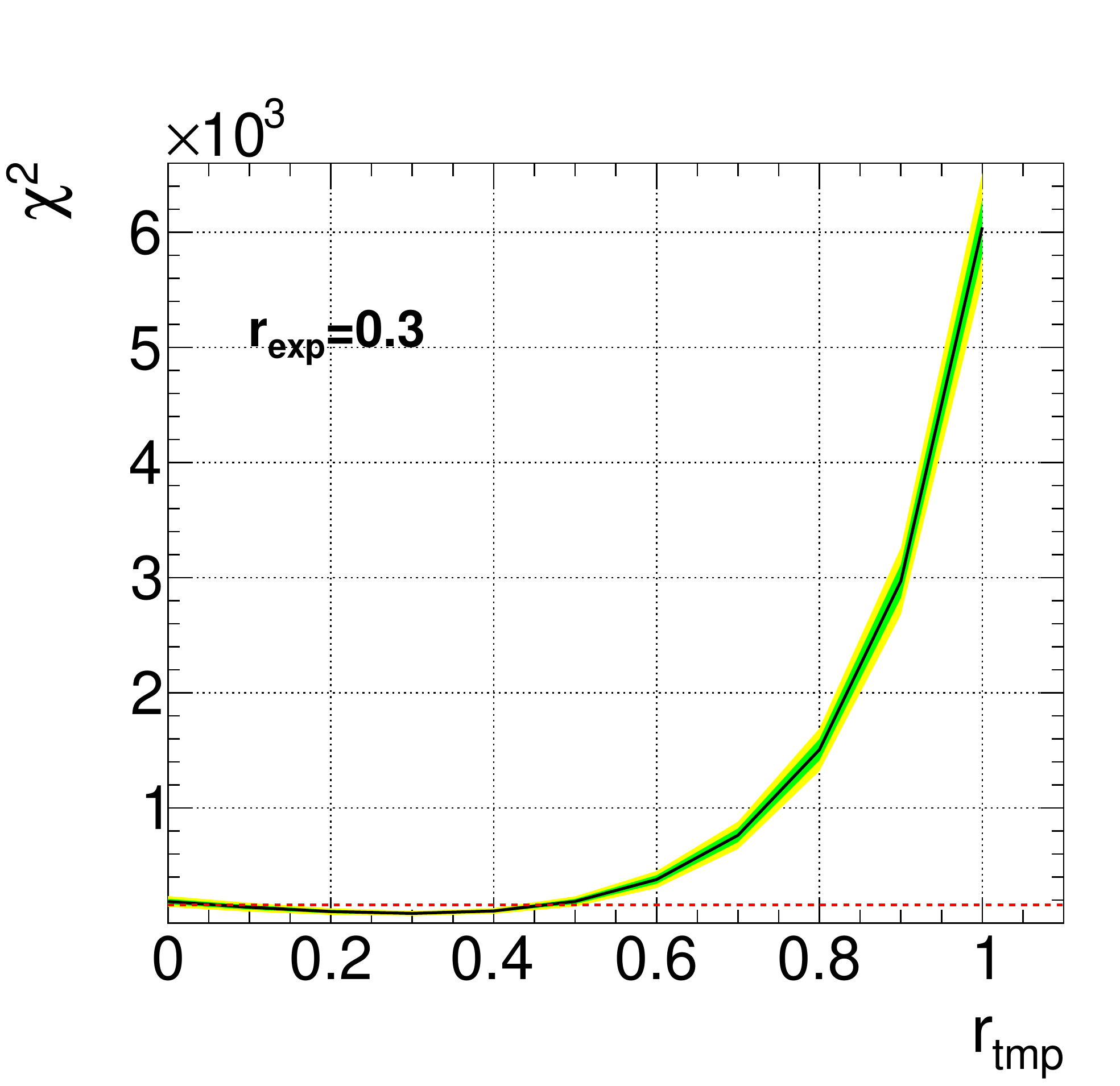}
 \caption{The chi-squared test of the angular histogram (left) and the energy-angular distribution (right), between the anisotropy of template $r_\mathrm{tmp}$ and that for the pseudo-experiment $r_\mathrm{exp}$. The target atom is F and the energy threshold $E_R^{\mathrm{thr}} = 20$ keV. In the pseudo-experiment 
the event number $5\times 10^3$ (left) and $6\times 10^3$ (right). The red dashed line represents 90\% CL. Green and yellow green bands correspond to 68 \% and 95 \% C.L., respectively.}
\label{fig:chi2_FERnon0}
\end{figure}

\begin{figure}[t!]
%\vspace{-0.5cm}
 \centering
 \includegraphics[width=5.5cm,clip]{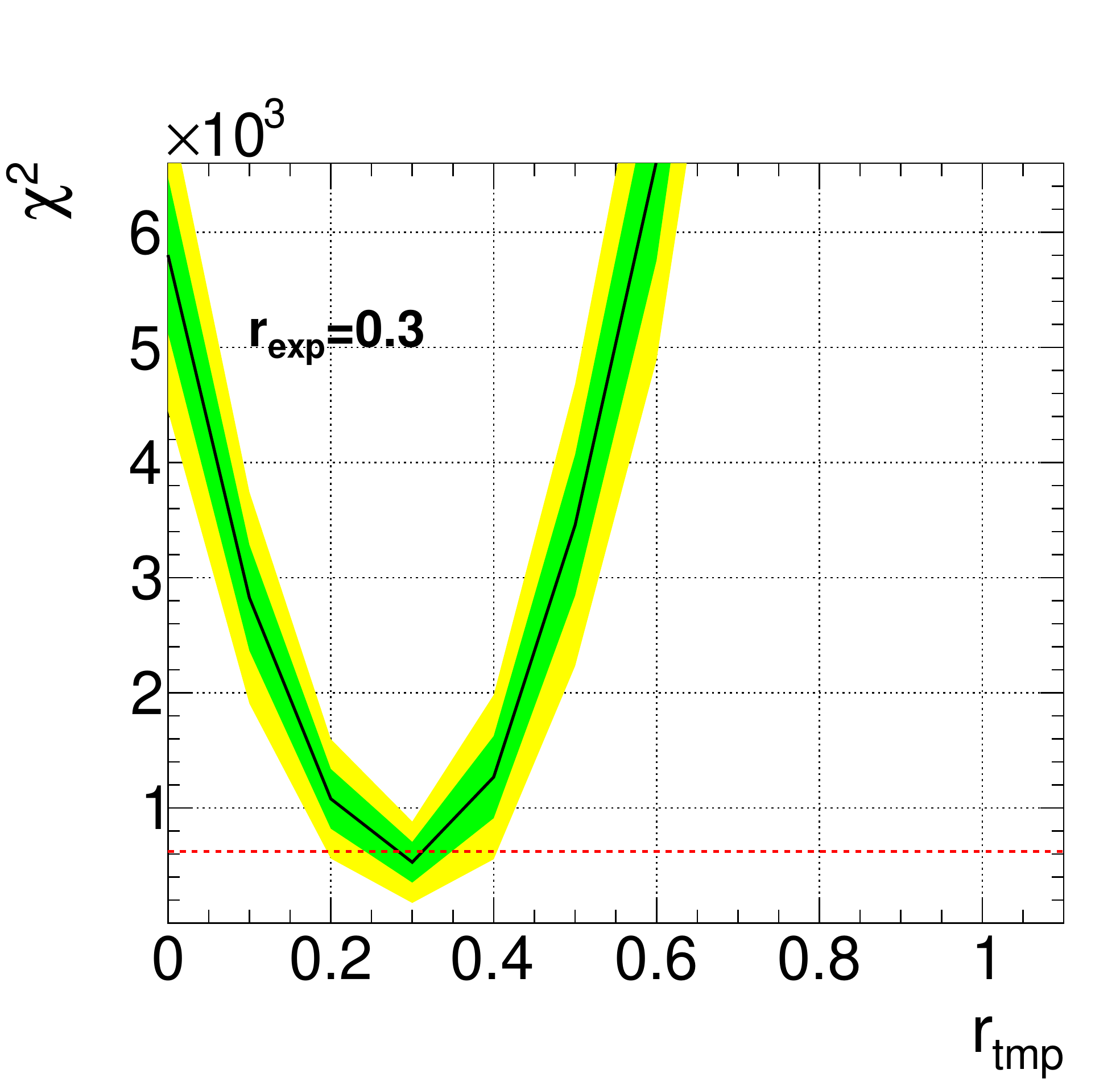}
  \includegraphics[width=5.5cm,clip]{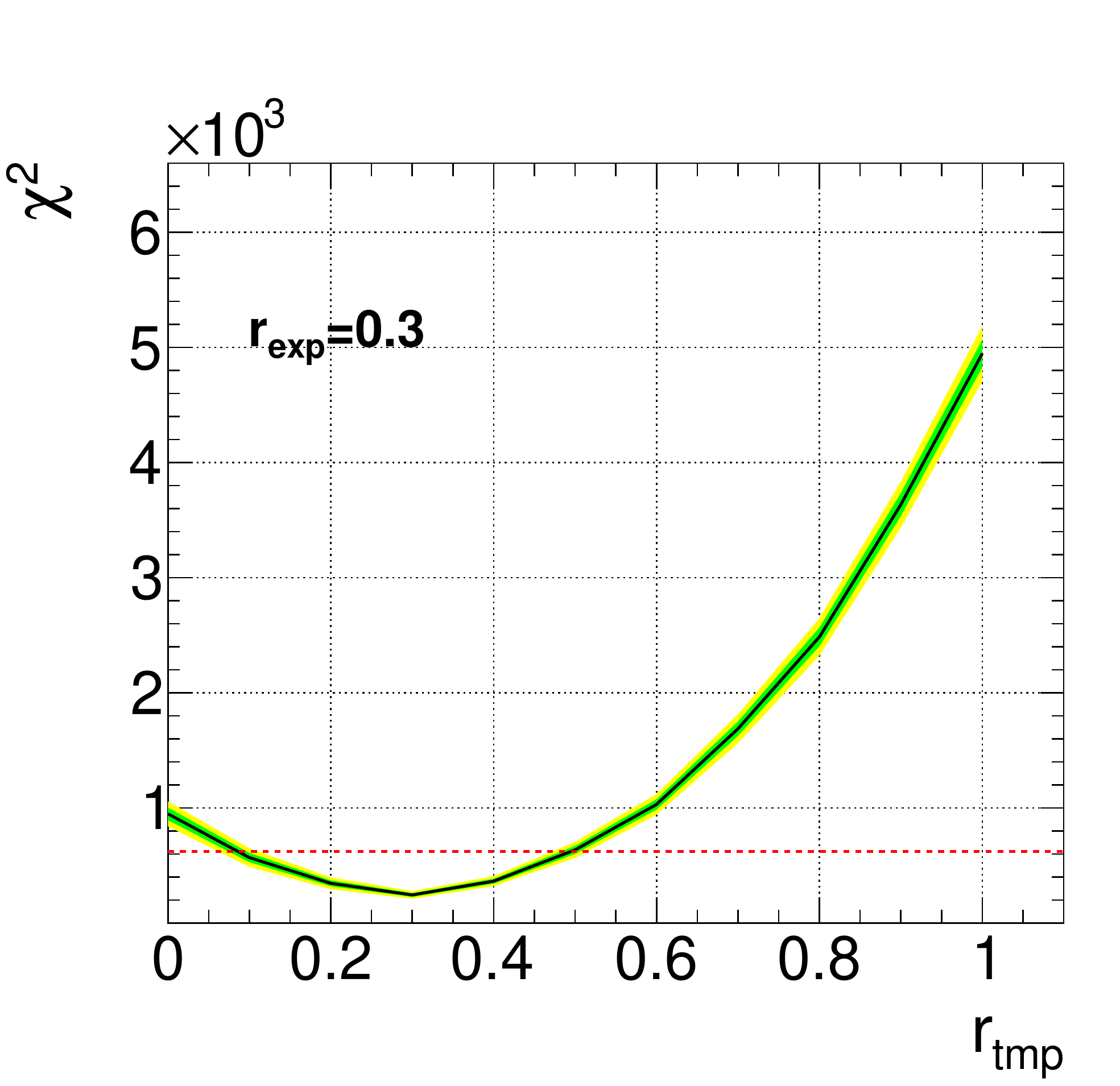}
 \caption{Legend is same as Fig. \ref{fig:chi2_FERnon0} for target Ag. The the energy threshold $E_R^{\mathrm{thr}} = 50$ keV. In the pseudo-experiment 
the event number $2\times 10^4$ (left) and $6\times 10^4$ (right).}
\label{fig:chi2_AgERnon0}
\end{figure}

If dark matter mass is suggested by other experiments such as indirect detection and collider experiment, the energy threshold of detector can be optimized to the mass. 
In Fig. \ref{fig:chi2_FERnon0} and \ref{fig:chi2_AgERnon0}, the chi-squared test for target F and Ag are shown, respectively. In the figures the energy threshold is optimized to the dark matter mass to minimize the required event number for the discrimination.
The parameter region is divided into $10 \times 20$ bins for the energy-angular distribution and 10 bins for the angular histogram to perform the chi-squared test. 
For simplicity, no background, perfect resolution of the detector and mass relation $m_\chi=3m_A$ are supposed for simplicity, in other words, $m_\chi=57$ GeV for target F and $324$ GeV for Ag.
As demonstrated in the figures, if the anisotropic distribution ($r=0.3$) is realized, O($10^3$) - O($10^4$) events are required to reject the anisotropic distribution by $90$\% confidence level (CL).  Though same mass relation is assumed, the required event number for target Ag is an order of magnitude larger than that for target F because the distribution of data is distorted due to nuclear form factor for heavy target case.


\subsection{Case 2 : Dark matter mass is not known.}
\begin{figure}[t]
 \centering
 \includegraphics[keepaspectratio, scale=0.2,clip]{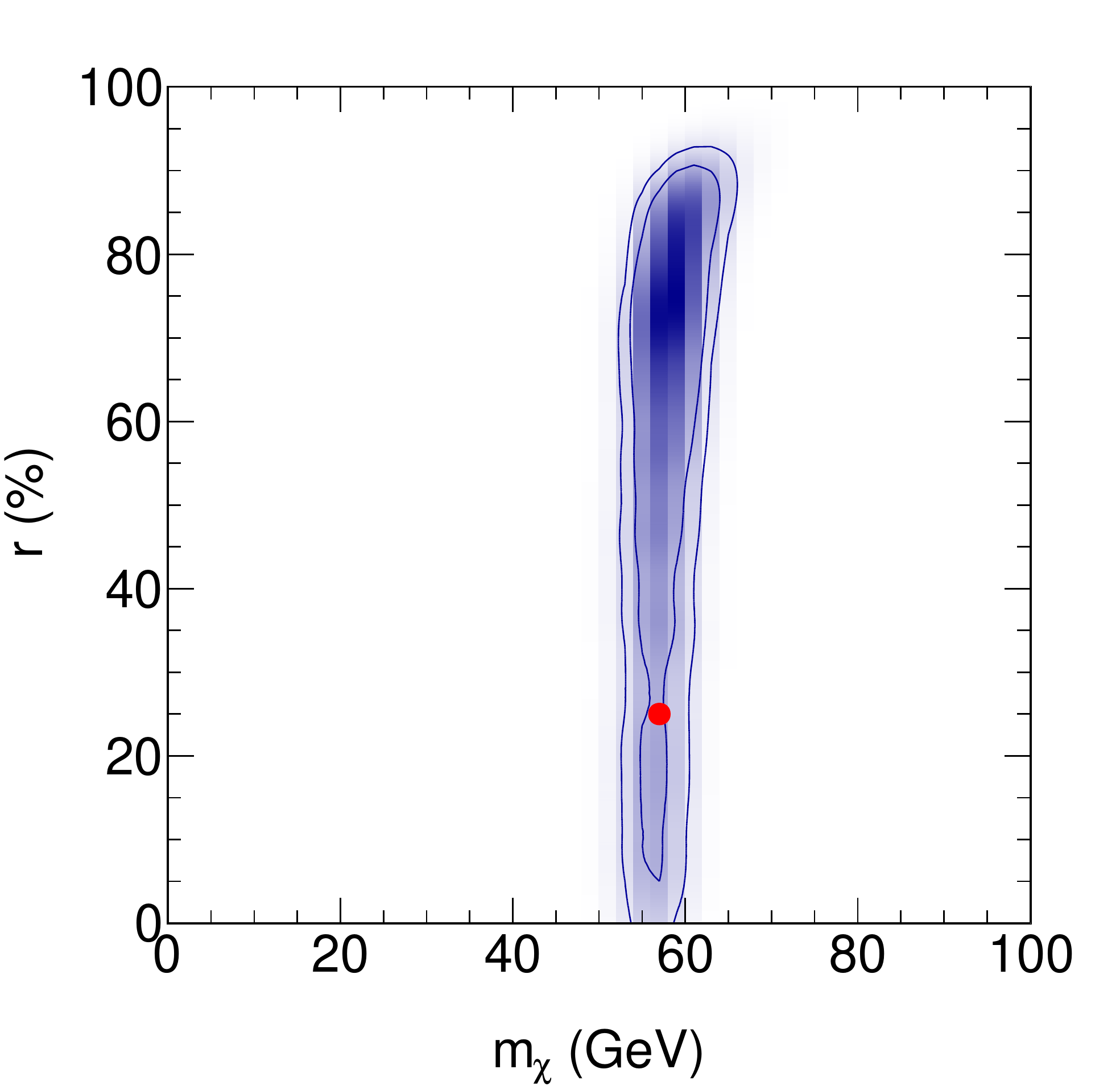}
\includegraphics[keepaspectratio, scale=0.2,clip]{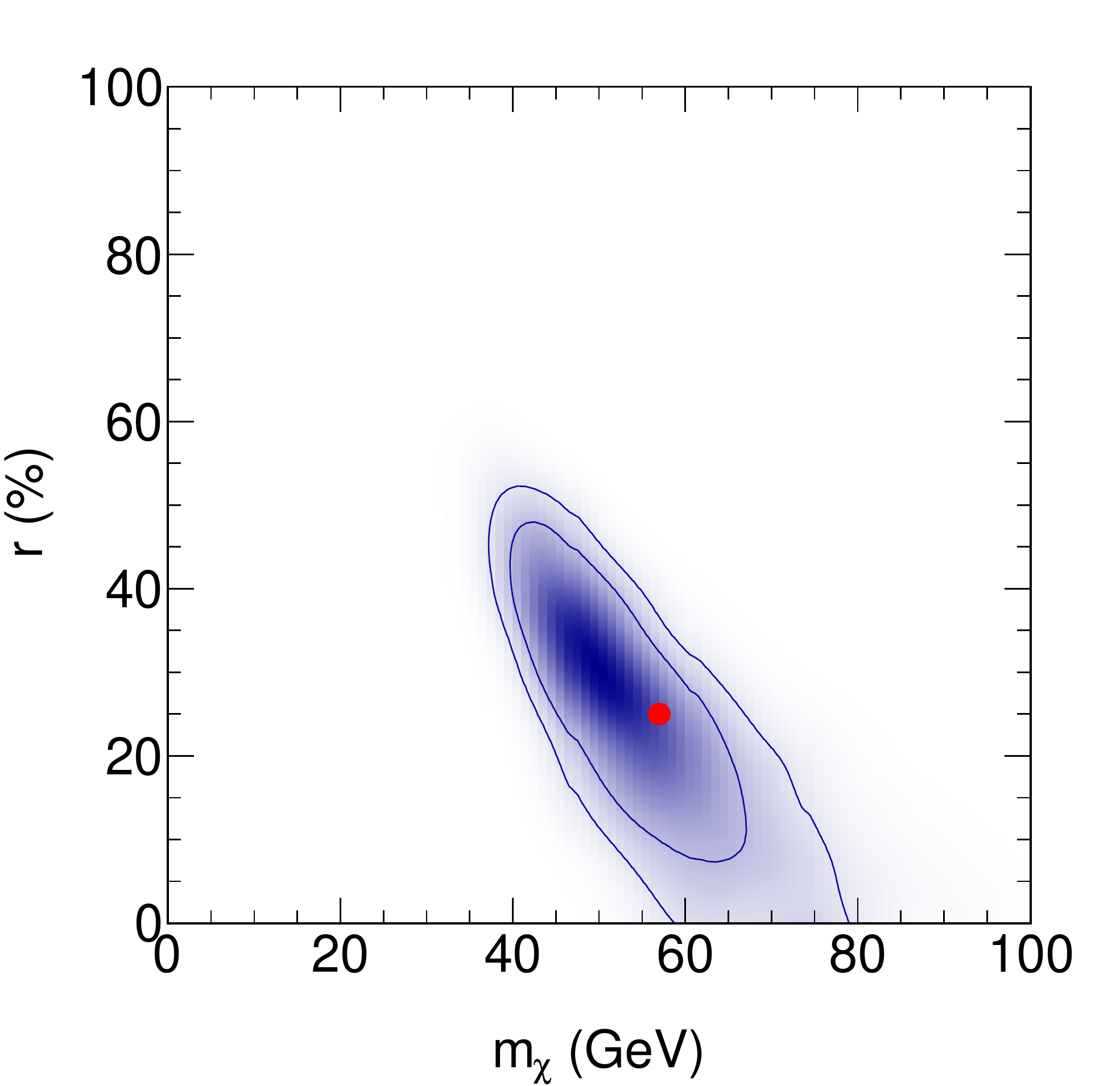}
\includegraphics[keepaspectratio, scale=0.2,clip]{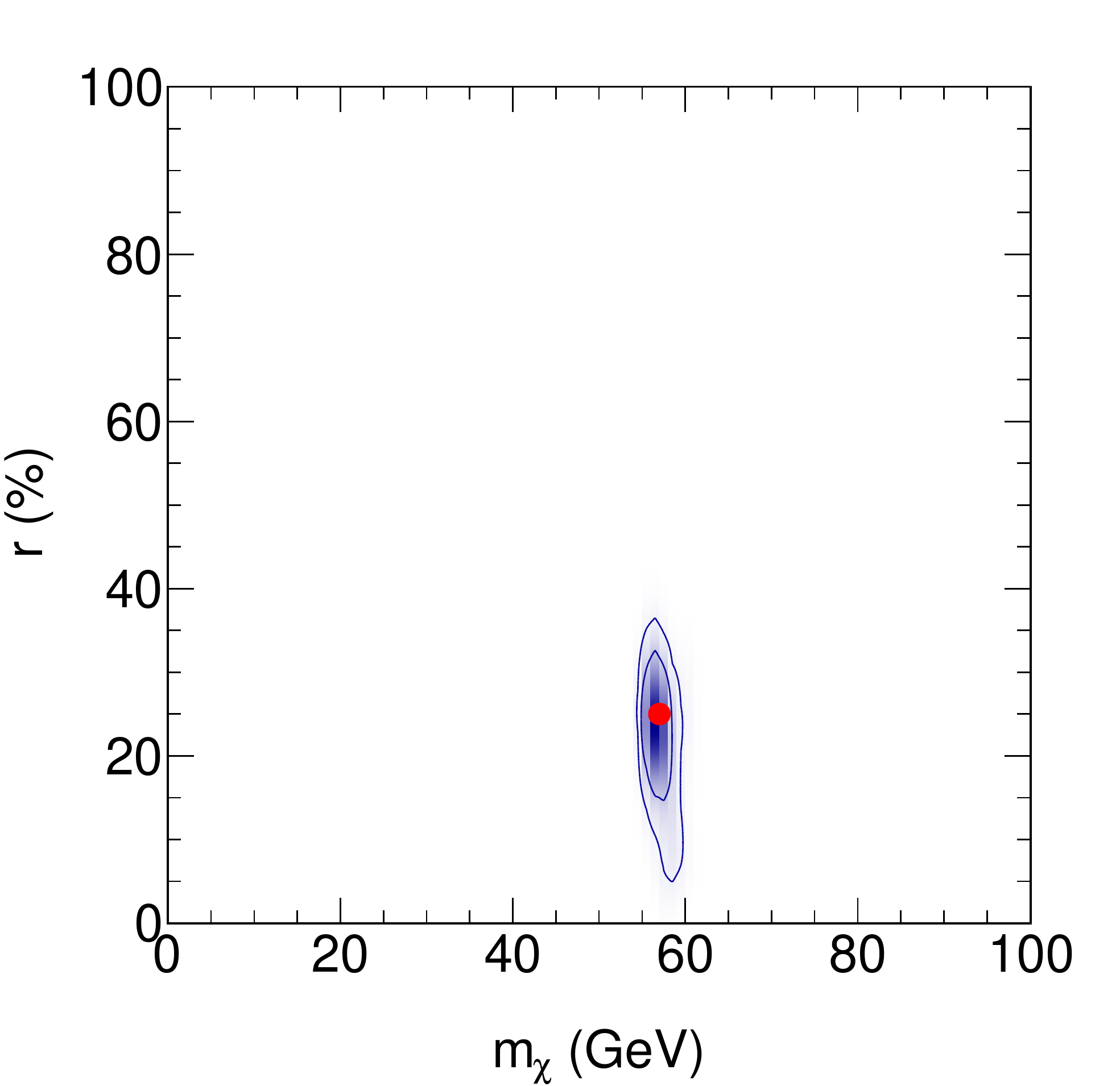}
 \caption{The 2D posterior probability distributions in the dark matter mass and anisotropy space for target F. Red points correspond to the input parameters in simulation. Inner and outer contours show 68\% and 90\% C.L, respectively. 
 Left: Only data of recoil energy $E_R$ is used. Center: Only data of scattering angle $\cos{\theta}$ is used. Right: Both recoil energy and scattering angle are used.}
 \label{fig:likelihoodF}
\end{figure}

\begin{figure}[t]
%\vspace{-0.5cm}
 \centering
 \includegraphics[keepaspectratio, scale=0.2,clip]{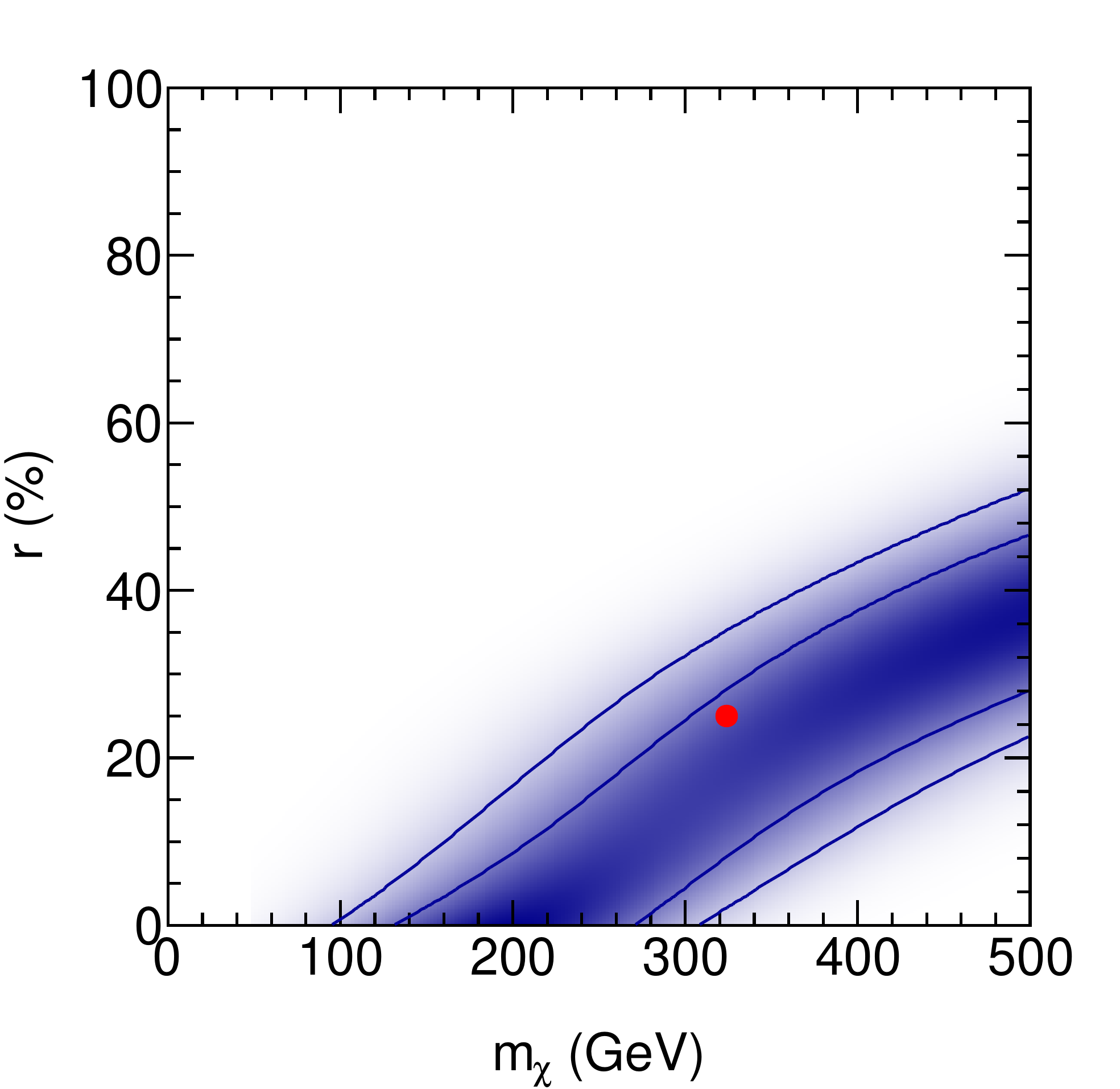}
\includegraphics[keepaspectratio, scale=0.2,clip]{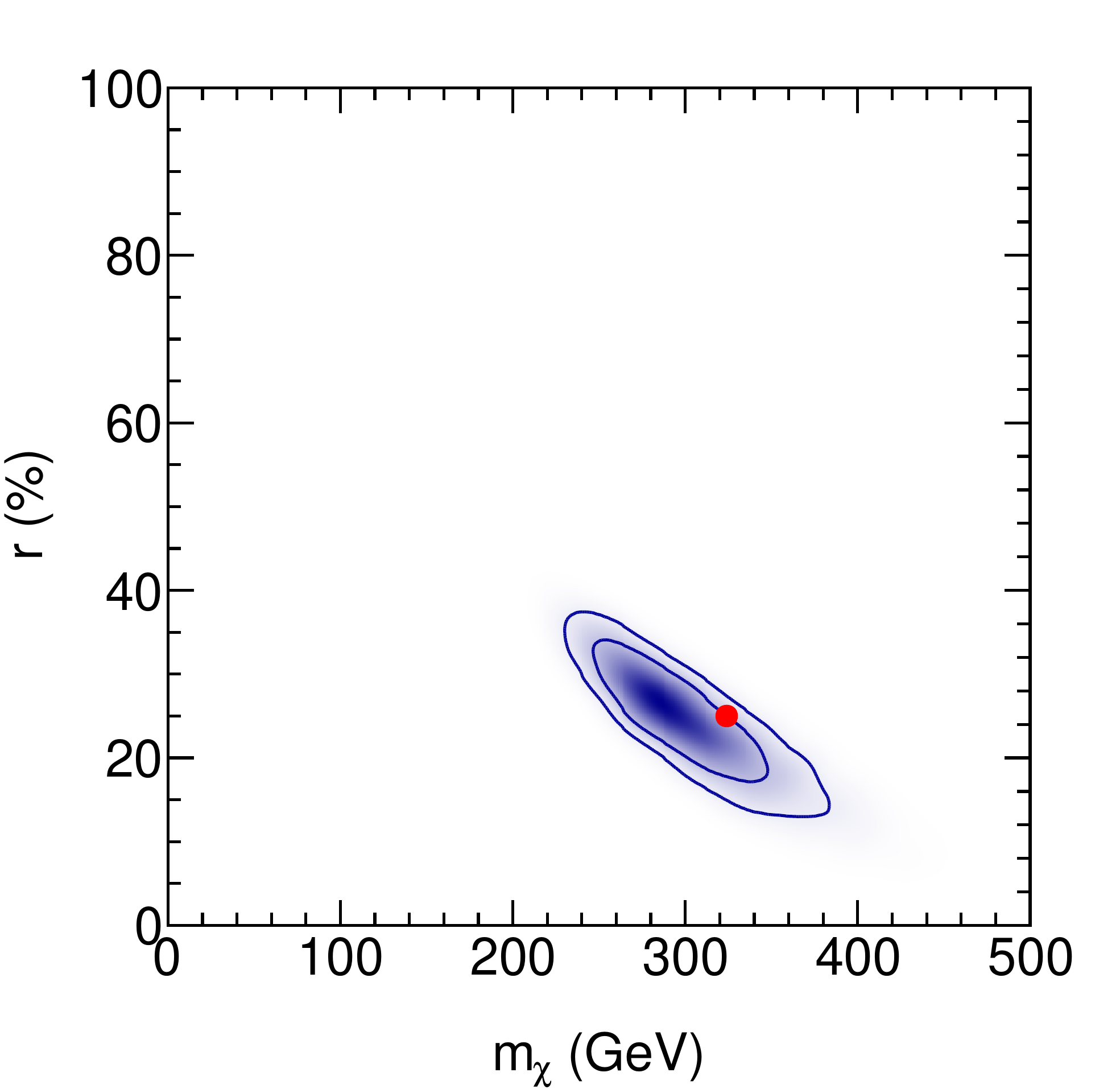}
\includegraphics[keepaspectratio, scale=0.2,clip]{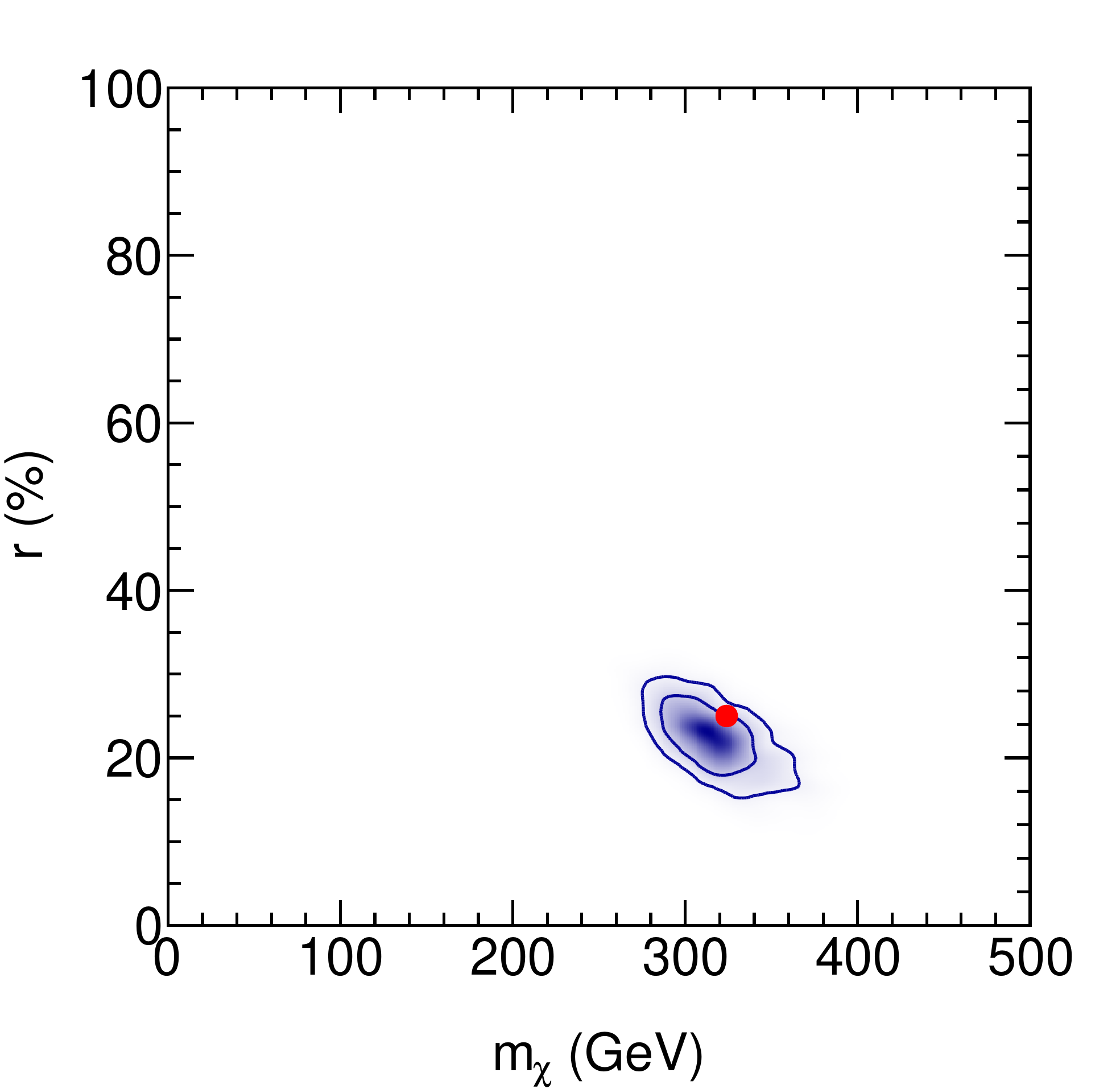}
 \caption{Legend is same as Fig. \ref{fig:likelihoodF} but for target Ag.}
 \label{fig:likelihoodAg}
\end{figure}

If the dark matter mass is not known, we should give constraint to two-dimensional parameter space of the anisotropy and the mass. In Fig. \ref{fig:likelihoodF} and Fig.\ref{fig:likelihoodAg}, the likelihood analysis of them for F and Ag are shown, respectively. 
A likelihood function is defined by
\begin{equation}
L=\prod_{i=1}^{N_{\rm{ene}}}\prod_{j=1}^{N_{\rm{ang}}} P(M_{ij}|\bar{M}_{ij}) \nonumber \ ,
\end{equation}
where $P(M_{ij}|\bar{M}_{ij})$ is the Poisson distribution function with the observed number of events $M_{ij}$ and the expected number of events $\bar{M}_{ij}$. The index of $i$ and $j$ represent the $i$-th energy bin and $j$-th angular bin. If we analyze with either the recoil energy or the scattering angle, the parameters are not suggested strictly or sometimes can be biased as seen in left panel of Fig. \ref{fig:likelihoodF}. Compared to them, the likelihood estimation works better in the case that both the recoil energy and the scattering angle are used in the analysis while in Sec. \ref{subsec:case1} chi-squared test using only recoil energy data can work better than the test using both the recoil energy and the scattering angle.

\section{Conclusion}
We investigate conditions to discriminate anisotropy of velocity distribution
by directional direct detection of dark matter. 
Supposing the isotropic and an anisotropic distributions,
 datasets of recoil energy and scattering angle of dark matter-target scattering 
in directional detection are generated in the Monte-Carlo simulation.
By analyzing the difference between ideal template and pseudo-experimental data, 
we showed that $O(10^3-10^4)$ events are required to reject the isotropic distribution 
if the anisotropic one is realized and the dark matter mass is known. 
Case that the dark matter mass is unknown is also studied. In that case
both the dark matter mass and the anisotropy should be discussed at the same time,
and likelihood analysis using both the recoil energy and the scattering angle can give better constraint
to the parameters than analysis using either of them.

%\newpage
\bibliographystyle{unsrt}

\end{document}